\documentclass[11pt,twoside]{article}

\usepackage{asp2004}
\usepackage{epsf}
\usepackage{psfig}
\usepackage{lscape}

\begin{document}
\title{Ionization structure in the disks and winds of B[e] stars} 
\author{Michaela Kraus \& Henny J.G.L.M. Lamers} 
\affil{Sterrekundig Instituut, Utrecht University, Princetonplein 5, 3584 CC Utrecht, The Netherlands} 

\begin{abstract} 
We investigate the ionization structure in the non-spherical winds and disks 
of B[e] stars. Especially the luminous B[e] supergiants seem to have 
outflowing disks which are neutral in hydrogen already close to the stellar 
surface. The existence of neutral material so close to the central star is
surprising and needs to be investigated in detail. We perform our
model calculations mainly in the equatorial plane, trying to find a plausible
scenario that leads to recombination in the vicinity of the hot stars.
Two different approaches are presented that both result in a hydrogen neutral 
equatorial region. We especially focus on the influence of stellar rotation 
which is known to play a significant role in shaping non-spherical winds. We
show that a rotating star can have a neutral equatorial wind, even without
the need of a density enhancement due to bi-stability or wind
compression, but simply due to graviational darkening in combination with
the known high mass-loss rates of the B[e] stars and especially the B[e] 
supergiants. 
\end{abstract}


\section{Introduction} 

The circumstellar material of B[e] stars is known to be non-spherically 
symmetric, and the best studied sample of B[e] stars, the B[e] supergiants, 
even show strong indication for the existence of an equatorial gas and dust 
disk. The hybrid character of their optical spectra have been interpreted
by a non-spherical wind with a typical line-driven wind in polar, and a slow,
high-density wind in equatorial direction \citep[and this volume]{Zick1985}. 
Indications for the
existence of a neutral gas and dust disk come (i) from molecular emission like the CO first
overtone bands \citep{mcgrega,mcgregb}, (ii) the existence of dust causing a strong mid-IR 
excess emission \citep[see e.g.][and references therein]{Voors}, and (iii) the 
detection of strong
[O{\sc i}] emission in the high-resolution optical spectra of several 
B[e] stars and supergiants \citep[and poster presentation, this 
volume]{KB2005,Ketal2005}, the latter even speaks in favour of a huge amount 
of hydrogen neutral material in the vicinity of these luminous stars.
This neutral material can only 
exist in the dense disks around these luminous stars, and model calculations 
have shown that these disks must be neutral already near the stellar surface 
to account for the observed line luminosities and to keep the needed disk mass 
loss rate at a reliable value \citep[and Kraus et al. poster presentation, this 
volume]{KB2005}. The goal of our study is therefore to find scenarios which 
allow neutral material to exist close to the surfaces of these stars.

\section{Ionization structure calculations in non-spherical winds}

We present two different approaches which both lead to the formation
of a neutral equatorial region. The first one is an empirical model
which assumes a non-spherical wind with a surface mass flux
that increases from pole to equator. The second one deals with the influence
of rigid stellar rotation.
 
\subsection{Empirical non-spherical wind model} 

In a first attempt we developped a code to calculate the ionization structure 
in a pure H plus He wind \citep{KL2003}. We used empirically defined 
latitude dependent mass flux and escape (or terminal) velocity distributions 
(left and mid panel of Fig.\,1), which define a surface density distribution 
that accounts for
the observed velocity and density ratios between
equator and poles, and creates a disk-like structure with an opening angle of 
$20\deg - 30\deg$ (mid and right panel of Fig.\,1).

\begin{figure}[t]
\plotone{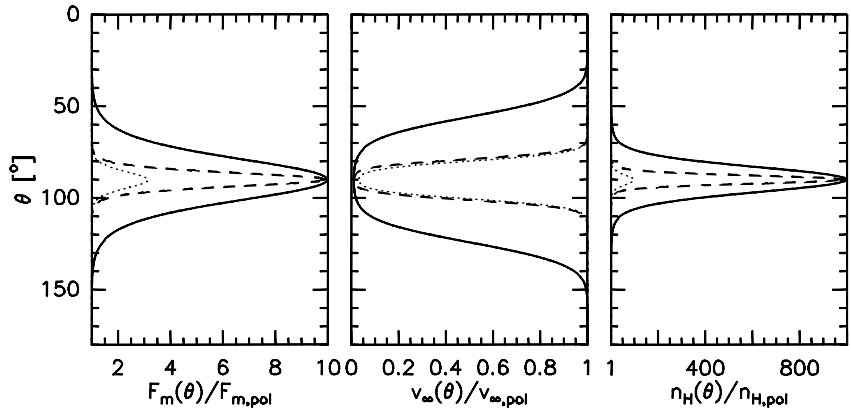}
\caption{Distribution of the mass flux (left), escape (and hence terminal, 
because $v_{\infty}\sim v_{\rm esc}$) 
velocity (mid) and hydrogen density (right) over the stellar surface. The 
parameters are normalized to their polar values. Different line styles are for
different model parameters as specified in \citet{KL2003}.}
\end{figure}

We solved the ionization balance equations for H and He in the on-the-spot
approximation which means that all photons generated via recombination
and suitable for re-ionization are absorbed in situ. The electron temperature
and radial wind velocity were set constant and we used the point source
approximation for the stellar radiation field. These restrictions resulted
in the calculation of an {\it upper limit} of the recombination distances
which are plotted for three different model parameters in Fig.\,2 (for more
model details see Kraus \& Lamers 2003).

\begin{figure}[ht]
\plotone{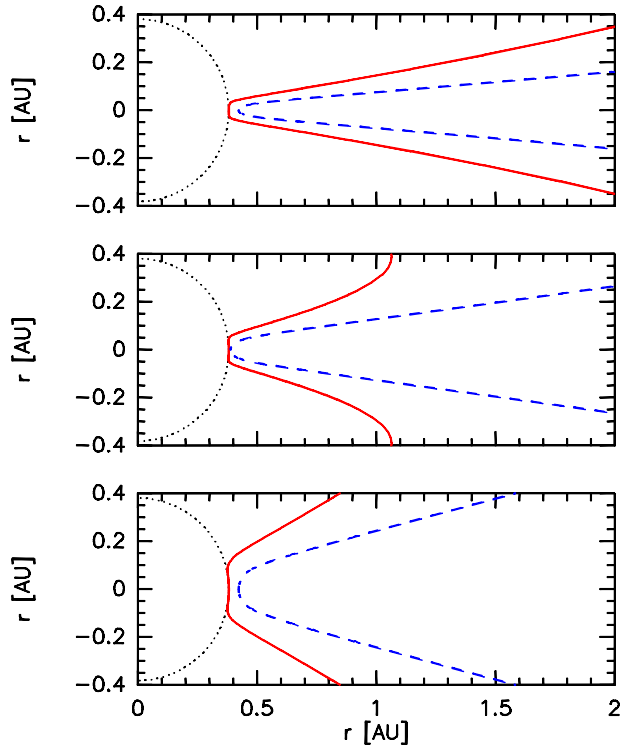}
\caption{Three examples for the ionization structure of H (dashed) and He
(solid) in non-spherical winds \citep[taken from][]{KL2003}. The material to
the right of the full or dashed line is neutral. Both elements
clearly recombine close to the stellar surface in the
equatorial plane, leading to a hydrogen neutral disk.}
\end{figure}

All these models for which the total mass loss rates were in the range 
$2.4\times 10^{-6}\ldots 2.4\times 10^{-5}\,$M$_{\odot}$yr$^{-1}$, which is
a reasonable range for B[e] supergiants, resulted in a hydrogen neutral disk
close to the hot stellar surface.

\subsection{Stellar rotation and the formation of a non-spherical wind}

\begin{figure}
\plotone{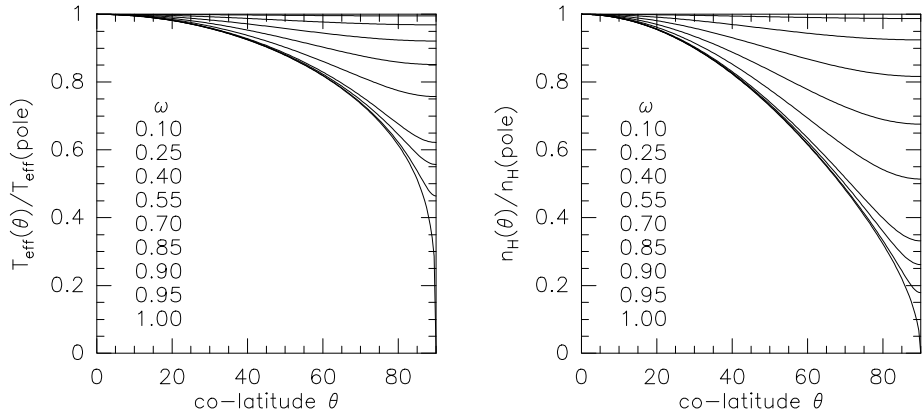}
\caption{Normalized surface distributions of the effective temperature (left) 
and the wind density (right) as a function of the co-latitute, $\theta$. The 
curves from top to bottom are for an increasing rotational velocity indicated by
the parameter $\omega$, which is the ratio of the equatorial rotation velocity 
over the critical velocity.}
\label{kraus_fig3}
\end{figure}

During the last years, the non-negligible influence of stellar rotation on
stellar evolution has been investigated in detail \citep[see e.g.][and this 
volume]{Maeder99,MaMe00,MeMa00}, and it has been suggested that the appearance 
of non-spherical winds around luminous stars might be caused by rotation 
\citep{Maeder02,MaDes}. We therefore started to investigate the influence of 
rigid stellar rotation on the formation of a hydrogen neutral disk. Rotation 
causes a flattening of the stellar surface and therefore a reduction of the
local net gravity in the equatorial region. The decrease in gravity from 
pole to equator equally results in a decrease of the stellar flux which
is proportional to the local net gravity. Hence, the effective temperature
also decreases from pole to equator (left panel of Fig.\,\ref{kraus_fig3}), 
known as the gravity darkening \citep[or polar brightening,][]{Zeipel}.

The latitude dependence of the gravity and effective temperature has also 
impacts on the stellar wind parameters: The escape velocity following from the
balance between gravitational and centripetal forces becomes latitude dependent,
decreasing from pole to equator. The same holds for the terminal wind velocity
which is (for line-driven winds) proportional to the escape velocity.
And even the mass flux from the star tends to decrease from pole to equator
when implementing gravity darkening into the CAK theory, as shown by
\citet{Owocki1998}. 
 
More important for the ionization structure calculations is the density in
the wind. According to the mass continuity equation the hydrogen density
scales as $n_{\rm H}\sim F_{\rm m}/(v\,r^{2})$. A proper elaboration
of this ratio, taking into account the rotational distortion of the 
stellar surface, resulted in surface density distributions of rotating
stars as shown in Fig.\,\ref{kraus_fig3} (right panel). From this plot it is 
clear, that the density at any given distance also {\it decreases} from pole 
to equator. A rotating star, when neglecting additional effects like 
bi-stability or wind compression, will therefore have a less dense wind in 
the equatorial region.

Both important parameters in the ionization balance equations, i.e. the 
effective surface temperature {\it and} the surface density, decrease from pole 
to equator. While the decrease in surface temperature tends to decrease the 
number of available photons suitable to ionize H and He, the decrease in
surface density reduces the optical depth along the line of sight
from a point in the wind to the star. Both effects are therefore counteracting
with respect to the location where recombination takes place: While a reduction
of ionizing photons will shift the recombination distance towards the star,
the reduction in optical depth along the direction to the star will shift it 
further outwards. The outcome of the ionization balance equations is therefore 
unpredictible and very sensitive to the choosen input parameters. This makes it
worth to investigate the ionization structure in the wind of a rotating star in 
more detail.

In our model calculations, the wind is assumed to be composed of H and He, 
and we solve the ionization balance equations in the equatorial plane, only. 
The radiation field from the star is calculated properly in 2D, which 
means that for the available radiation at every point in the equatorial wind 
we integrate the stellar radiation over the non-spherical surface and take into 
account the latitude variations of the individual parameters like effective 
temperature, net gravity, escape velocity and mass flux.
We make, however, several assumptions and simplifications for the calculations 
which all have in common that they shift the location of recombination to 
larger distances. This means that we are calculating an {\it upper limit} for 
the recombination distances\footnote{For details on the assumptions and their 
influence see \citet{KL2005}.}.

\begin{figure}
\plotone{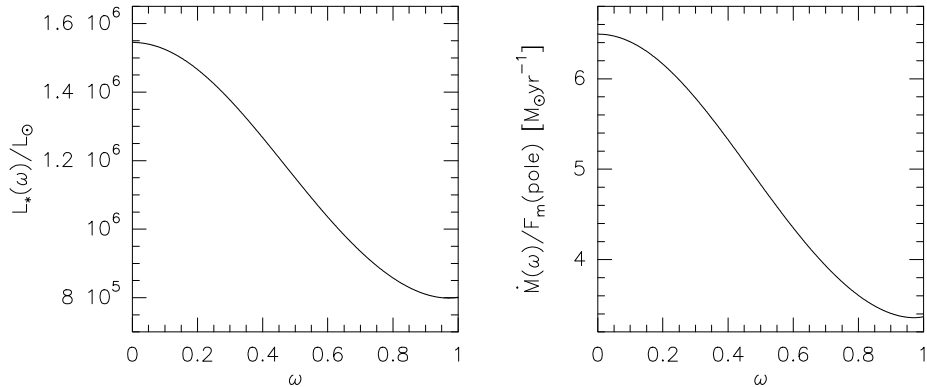}
\caption{Left: Stellar luminosity as a function of $\omega$ ($\omega = v_{\rm 
rot} / v_{\rm crit}$) for our model stars with $T_{\rm eff, pole} = 17\,000$\,K 
and $R_* = 82\,R_{\odot}$. Right: The ratio between the mass loss rate, 
integrated over the stellar surface, and the polar mass flux as a function of 
$\omega$. The decrease in both, luminosity and total mass loss rate, is about a 
factor of 2 between the non-rotating and the critically rotating star.}
\label{lum_mdot}
\end{figure}

We performed a series of test calculations where we fixed the polar values
of the stars, i.e. the polar effective temperature and gravity which determine
the radiation temperature of the ionizing stellar radiation, and the polar 
terminal velocity. Fixing the polar effective temperature means that we
are {\it not} calculating a star that is spinning up which would result in an 
increase of polar temperature with increasing rotation velocity, but we are
calculating stars with the same polar effective temperature having different
rotation velocities. This means that we are dealing with stars of {\it 
different luminosities}.

The difference in luminosity of these stars can be seen in Fig.\,\ref{lum_mdot} 
(left panel) where we plotted the distribution of the stellar luminosities of 
our model stars as a function of stellar rotation. This difference is largest 
between the non-rotating and the critically rotating star, and is about a 
factor 2.

Also shown in Fig.\,\ref{lum_mdot} (right panel) is the variation of the total 
mass loss of the rotating stars. For the choosen polar mass fluxes of $1.5
\times 10^{-6}$ and $1.5\times 10^{-5}$g\,s$^{-1}$cm$^{-2}$, the mass loss 
rates decrease from $10^{-5}$ and $10^{-4}$M$_{\odot}$yr$^{-1}$ (non-rotating) 
to $5\times 10^{-6}$ and $5\times 10^{-5}$M$_{\odot}$yr$^{-1}$ (critically 
rotating), respectively. 


\begin{figure}[ht]
\plotone{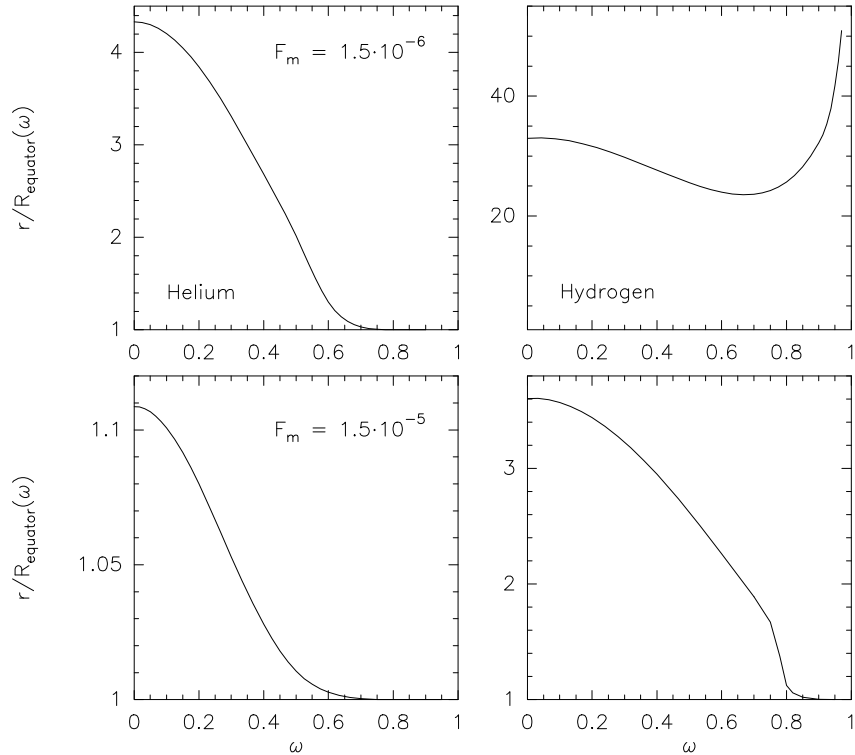}
\caption{Helium (left) and hydrogen (right) recombination distances in the 
equatorial plane as a function of $\omega$. The polar mass flux used for each 
model is indicated. While for the high mass flux model hydrogen recombines 
right above the stellar surface for $\omega \ge 0.8$, the recombination 
distance for the model with lower mass flux has a minimum at about $\omega 
\simeq 0.7$ and increases again for higher rotation rates.}
\label{ionstruc}
\end{figure}
                                                                                
\begin{figure}[ht]
\plotone{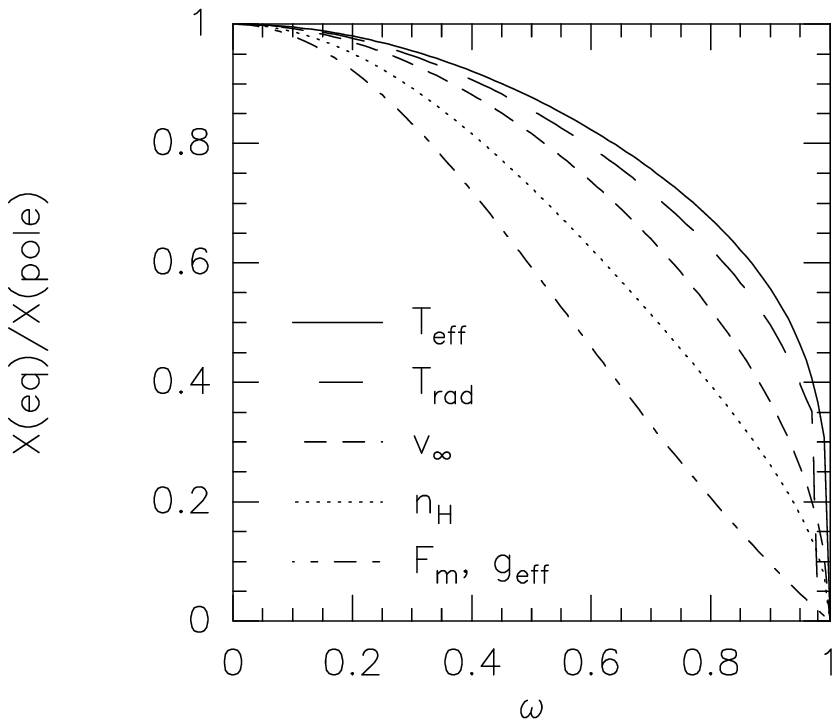}
\caption{Surface parameters in the equatorial plane as functions of the
rotation velocity. $n_{\rm H}$ is the density at any distance.}
\label{equatorial}
\end{figure}

First results of the ionization structure calculations in the equatorial winds
of luminous rotating stars are shown in Fig.\,\ref{ionstruc}. These 
calculations have been performed for polar mass fluxes that differ by a 
factor 10 to show clearly the influence of the input parameters.
For the high mass flux (bottom panels) both elements recombine right above the
stellar surface for $\omega \ge 0.8$, resulting in a (hydrogen) neutral
equatorial outflowing wind. The lower mass flux model (top panels) also shows
the trend of a decrease in recombination distance with increasing rotation. 
In the case of hydrogen, this distance has however a minimum around $\omega 
\simeq 0.7$, and increases again for higher rotation rates. This behaviour is 
unexpected but can be understood by looking at the variation of the individual 
equatorial surface parameters, especially of the temperature and wind density, 
with increasing rotation velocity as shown in Fig.\,\ref{equatorial}. This 
figure shows that with increasing rotation velocity, the density drops much 
quicker from pole to equator than the temperature. For recombination to take 
place right above the stellar surface, the number of ionizing photons has to be 
reduced. This can be done either by decreasing the radiation temperature, or by 
increasing the equatorial surface density and hence the optical depth. Since 
the decrease in radiation temperature is determined by the rotation velocity 
(with a fixed input polar value) we can only increase the input polar mass flux 
to achieve a higher surface density for a given rotation velocity. The density 
in the model shown in the top panel of Fig.\,\ref{ionstruc} is not high enough 
any more for stars with $\omega \ge 0.7$ to absorb the ionizing photons 
provided by the still rather high radiation temperature. Therefore, ionization 
takes over again and shifts the recombination distance for higher rotation 
velocities further out. Whether recombination takes place close to the star 
therefore sensitively depends on the choosen input parameters of the rotating 
star.

\section{Conclusions}

We presented two different wind scenarios that resulted in the formation
of a hydrogen neutral region in the equatorial plane: a non-spherical wind
model with a surface mass flux that increases from pole to equator, 
and a model investigating the influence of rigid rotation on the wind structure.
We show that the model of a rotating 
luminous star can have a hydrogen neutral equatorial region simply due to
the effects of gravitational darkening. In that case, it makes however no sense 
to speak of a "disk" because the density in the equatorial plane (when 
neglecting additional effects like bi-stability and/or wind compression)
is much lower than in the polar wind.

\acknowledgements 
M.K. acknowledges financial support from the Nederlandse Organisatie voor
      Wetenschappelijk Onderzoek grant No.\,614.000.310.


\vspace{-0.5cm}
\begin{thebibliography}{}
\bibitem[Kraus \& Borges Fernandes(2005)]{KB2005}
Kraus, M. \& Borges Fernandes, M. 2005, in The Nature and Evolution of Disks
Around Hot Stars, ed.\ R. Ignace \& K.G. Gayley, (San Francisco: ASP), p. 254
\bibitem[Kraus \& Lamers(2003)]{KL2003}
Kraus, M. \& Lamers, H.J.G.L.M. 2003, A\&A, 405, 165
\bibitem[Kraus \& Lamers(2005)]{KL2005}
Kraus, M. \& Lamers, H.J.G.L.M. 2005, {\sl submitted to A\&A}
\bibitem[Kraus et al.(2005)]{Ketal2005}
Kraus, M., Borges Fernandes, M., de Ara\'ujo, F.X. \& Lamers, H.J.G.L.M. 2005, 
A\&A, 441, 289
\bibitem[Maeder(1999)]{Maeder99}
Maeder, A. 1999, A\&A 347, 185
\bibitem[Maeder(2002)]{Maeder02}
Maeder, A. 2002, A\&A 392, 575
\bibitem[Maeder \& Desjaques(2001)]{MaDes}
Maeder, A. \& Desjaques, V. 2001, A\&A 372, L\,9
\bibitem[Maeder \& Meynet(2000)]{MaMe00}
Maeder, A. \& Meynet, G. 2000, A\&A 361, 159
\bibitem[McGregor et al.(1988a)]{mcgrega}
McGregor, P.J., Hyland, A.R. \& Hillier, D.J. 1988a, ApJ 324, 1071
\bibitem[McGregor et al.(1988b)]{mcgregb}
McGregor, P.J., Hillier, D.J. \& Hyland, A.R. 1988b, ApJ 334, 639
\bibitem[Meynet \& Maeder(2000)]{MeMa00}
Meynet, G. \& Maeder, A. 2000, A\&A 361, 101
\bibitem[Owocki et al.(1998)]{Owocki1998}
Owocki, S.P., Cranmer, S.R. \& Gayley, K.G. 1998, Ap\&SS 260, 149
\bibitem[Voors(1999)]{Voors}
Voors, R.H.M. 1999, PhD thesis, Utrecht University
\bibitem[von Zeipel(1924)]{Zeipel}
von Zeipel, H. 1924, MNRAS 84, 665
\bibitem[Zickgraf et al.(1985)]{Zick1985}
Zickgraf, F.-J., Wolf, B., Stahl, O., Leitherer, C. \& Klare, G. 1985, A\&A 143, 421
\end{thebibliography}
\end{document}